\documentclass[12pt,twocolumn]{iopart}
\usepackage[english]{babel}
\usepackage[applemac]{inputenc}
\usepackage{bbm}
\usepackage{iopams}
\usepackage{amsthm}
\usepackage[toc]{appendix}
\usepackage{color}

\newcommand{\qsp}[2]{\,\ensuremath{\raise.5ex\hbox{$#1$}\big\slash\raise-.5ex\hbox{$#2$}}}

\renewcommand{\Tr}[1]{\mathrm{Tr}\left[#1\right]}

\renewcommand{\d}{d} 
\newcommand{\dl}{\mathsf{d}_\ell} 
\renewcommand{\H}{\mathcal{H}}

\renewcommand{\P}{\mathcal{P}}

\newcommand{\Ad}{\mathsf{Ad}}

\newcommand{\cad}{\mathsf{ad}^*}

\renewcommand{\S}{\mathcal{S}}

\newcommand{\pure}[1]{
\left| #1 \right>\left< #1\right|
}
\newcommand{\ket}[1]{
\left| #1 \right>
}
\newcommand{\bra}[1]{
\left< #1 \right|
}

\newcommand{\scalar}[2]{
\left< #1 | #2 \right>
}
\newtheorem{definizione}{Definition}[section]

\newtheoremstyle{Rem}
{}
{}
{}
{}
{\itshape}
{:}
{\newline}
{}
\theoremstyle{Rem}
\newtheorem*{Remark}{\textsf{Remark}}

\theoremstyle{classic}

\begin{document}
\title{Geometry of mixed states for  a q-bit and  the quantum Fisher information tensor}
\author{E Ercolessi$^1$, M Schiavina$^2$}
\address{$^1$ Dipartimento di Fisica, Università di Bologna and INFN, Sezione di Bologna, via Irnerio 46, 40127 Bologna, IT}
\address{$^2$ Institut f\"ur Mathematik, UZH, Winterthurerstrasse 190, Zürich, CH}
\ead{ercolessi@bo.infn.it}
\begin{abstract}
After a review of the pure state case, we discuss from a geometrical point of view the meaning of the quantum Fisher metric in the case of mixed states for a two-level system, i.e. for a q-bit, by examining the structure of the fiber bundle of states, whose base space can be identified with a co-adjoint orbit of the unitary group. We show that the Fisher Information metric coincides with the one induced by the metric of the manifold of $SU(2)$, i.e. the 3-dimensional sphere $S^3$, when the mixing coefficients are varied. We define the notion of Fisher Tensor and show that its anti-symmetric part is intrinsically related to the Kostant Kirillov Souriau symplectic form that is naturally defined on co-adjoint orbits, while the symmetric part is nontrivially again represented by the Fubini Study metric on the space of mixed states, weighted by the mixing coefficients.
\end{abstract}

\section*{Introduction} \label{se:intro}    

The space of quantum states carries a very rich geometric structure and its topological as well as differential properties play an important role in the description of quantum mechanical systems. Intrinsic properties are indeed helpful in the prediction of quantum behaviour and in the treatment of the semiclassical limit (see \cite{erc} for a review). A well known example of this influence is given by  Berry's phase, originally introduced to describe cyclic unitary evolution of pure states \cite{berry} and then put in a geometrical context and generalized in various directions \cite{ahran1, sam}, also to the case of mixed states for open systems (see for example \cite{rez}). In recent times geometric structures of quantum states have been exploited also in the framework of quantum information  theory \cite{beng} and entanglement measures \cite{brohug}.
Within this context, these intrinsic features appear to have consequences on the possibility of extracting maximum information and attaining optimal bounds. 

In this set-up, one of the indices that have been studied most extensively is  Fisher information. The relationship between quantum statistics and optimization of the classical Fisher information for pure and non-degenerate mixed states has been clarified in the seminal papers \cite{brau,cav}, solving the problem in an algebraic setting and addressing the connection between the optimization problem and generalized Heisenberg relations. Important recent developments are carried on in \cite{barn} for two level pure states systems and in \cite{luati2} for two level mixed states systems. Furthermore, this link has been re-examined from different points of view \cite{kol} as well as applied with success to quantum metrology (see \cite{esc} and references therein).

As back as in 1945, it was first Rao \cite{rao} who related the geometry of the parameter-space submanifold in the the Hilbert space to the notion of  classical Fisher information, which is indeed also called  ``Fisher-Rao metric". Further developments were presented in \cite{bur}. Geometrical tools to interpret statistical inference and state estimation problems in quantum mechanics have been recently proposed in some interesting papers \cite{bro, hug, hug2}. An analysis of the connection between Fisher indices and geometrical framework has been presented in a recent paper \cite{marmo}. There it is remarked that, since for pure states the quantum Fisher information metric is nothing but the Fubini-Study metric (see for instance \cite{ahran2, brohug} on this), the conditions for the attainability of the information bound can be read as geometric conditions, opening the way for subsequent generalizations. In this paper we would like to extend this result to the case of mixed states for a q-bit, following mostly \cite{chat} and some ideas from the construction presented in \cite{zyc}. Our aim is to work in a purely geometrical context. We will start by showing how to interpret and recover the quantum Fisher metric from a purely intrinsic geometric description of the space of quantum states, by looking at the fiber bundle structure of the latter. This  will allow us to study the properties of the full Fisher tensor, which includes also an antisymmetric part.

In sect. \ref{se:pure} we will first review the known links  between classical and quantum Fisher information for pure states, by also considering the optimization problem. Following \cite{marmo}, we  will examine these results from a geometrical point of view, by looking at the structures that one can build on the space of pure states of a $d$-level quantum system,  considering the fiber bundle structure originating when going from the Hilbert space to the space of rays.  In particular, we will show how this structures arise from the fact that we are working on a  co-adjoint orbit $\mathsf{U}(d)/(\mathsf{U}(1)\times \mathsf{U}(d-1))$ of the group $\mathsf{U}(d)$. 

We will move to consider mixed states for a q-bit in sect. \ref{se:mix}, where we will first describe how to calculate the classical and quantum Fisher indices in this case, via the so called symmetric logarithmic derivative. Then, we  will discuss the geometry of the space of states, which is now related to that of the group $\mathsf{U}(2)$ and of the co-adjoint orbit $\mathsf{U}(2)/(\mathsf{U}(1)\times\mathsf{U}(1))$. 

Finally, in sect. \ref{se:geofisher} we will define and study the full quantum Fisher tensor, composed by both the Fisher metric part and an anti-symmetric contribution, which we will identify with the Kostant Kirillov Souriau symplectic form that one can define on co-adjoint orbits \cite{kir}.

We will draw conclusions and look at further developments in section \ref{se:concl}.

\section{Pure states for q-dits} \label{se:pure}

In this section we will first review some well known results on the relationship between classical and quantum Fisher information indices and their use in optimization problems in quantum mechanics. Then we will give a geometrical interpretation of these results, by discussing the structure of the space of quantum states.

\subsection{Fisher information and the optimization problem} \label{se:fisherpure} 

Let us start by considering a generic quantum state described by a (pure or mixed) density operator $\rho(\theta)$, that may depend on a real parameter $\theta$. Given a measurable space $\cal{X}$ with measure $dx$, which represents the space of possible outcomes of the measure, we can define a  POVM by means of a set of non-negative and self-adjoint operators  $m(x)$ with $x\in \cal{X}$ satisfying:
\begin{equation}
\int_\mathcal{X} m(x) dx = \mathbb I \; , \label{ident}
\end{equation}
 if we assume that the outcome of a measurement on $\rho$ is a random  variable $X$ taking values in $\mathcal{X}$ and such that for any measurable subset ${\cal A} \in \cal{X}$ we have: $ Pr\{X\in {\cal A} \} = tr [ \rho M({\cal A})] $ with $M({\cal A}) = \int_{\cal{A}} m(x) dx$. This means that the outcome $X$ of  a measure on $\rho(\theta)$ is described by a probability density given by:
\begin{equation}
p(x; \theta) = \Tr{\rho(\theta) m(x)}  \; . \label{density}
\end{equation}
We will denote with ${\cal X}_+$ the complement in  ${\cal X}$ of the subset ${\cal X}_0 = \{ x \, : \, p(x ; \theta) = 0 \}$.\\
We can now form the {\it score function} $\ell_\theta$:
\begin{equation}
\ell_\theta = \log p(x; \theta)  \label{score}
\end{equation}
and, by denoting with $d\equiv d_\theta$ the derivative with respect to the parameter $\theta$, we define the {\it  classical Fisher information} to be
the expected value of the square of the derivative of the score function, namely:
\begin{equation}
\imath (\theta,m) \equiv  E((d\ell_\theta)^2) = \int_{\cal X} (d\ell_\theta)^2 p(x;\theta) dx = \int_{\cal X} \frac{\large(  \Tr{d\rho(\theta) m(x)}\large)^2}{\Tr{\rho(\theta)m(x)}} \, dx  \; . \label{fisher1}
\end{equation}
Here and in the following, unless explicitly stated otherwise, we will use the symbol $d$ to denote the differential, which is usually called derivative in literature, along the curve parametrized by $\theta$. We will see the importance of using differential forms rather than derivatives when turning to the geometric interpretation, later on.

If we define the {\it symmetric logarithmic derivative} $\dl \rho$ via:
\begin{equation}
d\rho = \frac{\rho \, \dl\rho + \dl\rho \, \rho}{2} \; , \label{logder}
\end{equation}
by using the self-adjointness of both $\rho$ and $m$, we find that the classical Fisher information is obtainable as:
\begin{equation}
\imath (\theta,m)  = \int_{\cal X_+} \frac{\large( \mathrm{Re}\  \Tr{\rho(\theta) \dl\rho(\theta) m(x)}\large)^2}{\Tr{\rho(\theta)m(x)}}  \, dx \; . \label{fisher2}
\end{equation}
One can also define the {\it quantum Fisher information} as the expected value of the logarithmic derivative of the density operator:
\begin{equation}
I_\theta \equiv E((\dl \rho)^2) = \Tr{\rho(\theta) (\dl \rho(\theta))^2} \; , \label{qfisher}
\end{equation}
which yields an upper bound for the classical Fisher information (\ref{fisher2}). This result \cite{brau, hel} follows from the following chain of inequalities (to simplify the notation we drop the $\theta$ and $x$ dependence inside the integrals):
\begin{eqnarray}
&& \imath(\theta,m) =  \int_{\cal X_+} \frac{\large( \mathrm{Re}\  \Tr{\rho \dl\rho  m }\large)^2}{\Tr{\rho m}}\, dx \nonumber \\
&& \leq \int_{\cal X_+} \frac{|  \Tr{\rho \dl\rho  m }|^2}{\Tr{\rho m}}\, dx
= \int_{\chi_+}  \frac{ | \Tr{ (m \rho )^{1/2})^\dagger (m^{1/2} \dl\rho  \rho^{1/2} )} |^2}{\Tr{\rho m}} \, dx \label{chain1}\\
 & &\leq  \int_{\chi_+} \Tr{ (m^{1/2}  \dl\rho  \rho^{1/2} )^\dagger (m^{1/2}  \dl\rho  \rho^{1/2} )   }  \, dx = 
    \int_{\chi_+} \Tr{m \dl\rho  \rho  \dl\rho   } \, dx \label{chain2}\\
  & &\leq  \int_{\chi}  \Tr{ m  \dl\rho  \rho  \dl\rho     } \, dx
   = \Tr{ \rho   (\dl\rho)^2   } = I_\theta \, \label{chain3}
\end{eqnarray}
where the first two equalities together hold iff 
\begin{equation}
m^{1/2} (\dl\rho)  \rho^{1/2}  \propto_\mathbb R m^{1/2}    \rho^{1/2} \; , \label{condition}
\end{equation}
while the last one is true iff $ \int_{\chi_0}  \Tr{ m  \dl\rho  \rho  \dl\rho     }=0$, where we used the symbol $\propto_{\mathbbm{R}}$ to mean proportional by means of a real number. Clearly a sufficient condition for this to hold is that $p(x,\theta)>0$ for almost all $x$, a fact that will always be assumed in the rest of this paper. 
 
The problem of finding a class of measurements that attain the quantum information bound is clearly of utmost importance in information theory and quantum measurement problems. An answer for pure states has been discussed by several authors \cite{brau,barn,luati1} by means of  algebraic tools. 

In the rest of this section we will not consider the most general situation. Aiming at giving a geometric interpretation of the algebraic result that we are going to explicitly show following \cite{barn}, we will  suppose to work in a finite dimensional Hilbert space ${\cal H} = \mathbb C^d$, thus describing a d-level quantum system or a q-dit (the  d=2 case representing a q-bit). We will also restrict our attention to projective measures, for which the $m(x)$ operators all have rank-one and can therefore be written as: 
\begin{equation}
m(x) = |\xi(x) \rangle \langle \xi(x) | \, . \label{xi}
\end{equation}
Moreover, one may look at measurements that attain the bound uniformly in the parameter $\theta$. 
Under these assumptions, for each value of $\theta$ we may choose an orthonormal basis $\{|i\rangle \}_{i=1}^n$ in ${\cal H} $ (dependent on the parameter $\theta$ that we omit to simplify the notation) such that:
\begin{equation}
\rho = |\psi \rangle \langle \psi | \;\;  \mbox{ with } \;\; |\psi\rangle = |1\rangle \; , 
\end{equation}
\begin{equation}
 |d\psi \rangle = \sum_{i=1}^d  a_i  |i\rangle \; , 
\end{equation}
\begin{equation}
m =  |\xi\rangle \langle \xi|  \;\;  \mbox{ with } \;\; |\xi\rangle = \sum_{i=1}^d  \xi_i  |i\rangle \; ,
\end{equation}
where the coefficients $ \xi_i =  \xi_i (x)$  must satisfy (\ref{ident}) or equivalently:
\begin{equation}
\int_{\cal X} \xi_i(x)^* \, \xi_j(x) = \delta_{ij}  \; . \label{delta} 
\end{equation}
We notice also that, since $\langle \psi | \psi \rangle =1$, the coefficient $a_1 = \langle \psi | d\psi\rangle$ is pure imaginary: $a_1 = i a, \; a\in \mathbb R$.\\
It is straightforward to verify that, for the pure case we are considering, the symmetric logarithmic derivative coincides with ordinary derivative, up to a factor 2:
\begin{equation}
\dl\rho = 2 d\rho \;. \label{dd}
\end{equation}
Thus, we have:
\begin{equation}
\imath(\theta,m) =  4 \int_{\cal X} \frac{\large( \mathrm{Re}\ [\xi_1(x)^* \, \sum_{i=2}^d \xi_i(x) a_i^*])^2}{|\xi_1|^2}\, dx  \; , 
\end{equation}
while
\begin{equation}
I_\theta  = 4 \sum_{i=2}^d | a_i|^2 \; . 
\end{equation}
Condition (\ref{condition}) translates into the following requirement:
\begin{equation}
\xi_1 \propto_{\mathbbm{R}}  \sum_{i=2}^d \xi_i(x) a_i^* \label{reach} \; , 
\end{equation}
which allows to reach the bound if we use also (\ref{delta}).

What we have obtained here, is a generalization of the proof to the d-level case of the derivation for the d=2 example given in \cite{barn}. 

We also remark that, if we choose the measure $m$ to be represented by one-dimensional operators depending on a continuous variable $x\in X$\footnote[7]{For example, $|\xi(x)\rangle$ may represents  coordinate  or coherent states.} we can set, as it is done in  \cite{marmo}:
\begin{equation}
\psi(x;\theta) = p(x;\theta)^{1/2} e^{i\alpha(x;\theta)} \; , 
\end{equation}
where $\psi(x;\theta)= \langle x| \psi(\theta)\rangle  $ is the wave function representing the state $|\psi\rangle$ in the $x$-representation:
\begin{equation}
|\psi(\theta)\rangle = \mathbb I |\psi(\theta)\rangle = \int_Xdx \,  |x\rangle \langle x| \psi(\theta)\rangle  \; . 
\end{equation}
It is not difficult to check that now the classical Fisher information index is simply given by:
\begin{equation}
\imath(\theta,m) =  \int_Xdx \, p(x;\theta)\, (d\log p(x;\theta))^2 \; , 
\end{equation}
while the quantum Fisher index is:
\begin{equation}
\fl I_\theta =  \int_Xdx \, p(x;\theta)\, (d\log p(x;\theta) )^2 +  \int_Xdx \, p(x;\theta)\, (d\alpha(x;\theta) )^2  - \left( \int_Xdx \, p\, d\alpha(x;\theta) \right) ^2\; . 
\end{equation}
Moreover, since $|d\psi\rangle  = \int_Xdx \, \left( d\sqrt{p} \, e^{i\alpha} + i d\alpha \sqrt{p} \right)$, condition (\ref{reach}) translates into $d\alpha = 0$. This makes the link between \cite{barn} and \cite{marmo} explicit. \\

\subsection{Geometric interpretation of Fisher information}   \label{interpure} 

It is a fairly acknowledged fact \cite{ahran2, brohug, kib} that a pure state of a quantum system is not simply described by a vector $|\psi\rangle$ in a Hilbert space ${\cal H}$. Rather it is best represented by a ray, since we have first to normalize vectors and then to remove phase redundancies. For a finite-dimensional Hilbert space ${\cal H} = \mathbb C^d$, the space of pure states may therefore be identified with the base space of a double principal fibration:
\begin{equation}
\begin{array}
[c]{rc}%
\mathbb{R}_{+} \hookrightarrow & \mathcal{H}_{0}=\mathbb C^d\backslash\left\{
\mathbf{0}\right\} \\
  & \downarrow\\
\mathsf{U}\left(  1\right)   \hookrightarrow & \mathbb{S}^{2d-1}\\
  & \downarrow\\
  & P\mathcal{H} \sim \mathbb C \mathbb P^{d-1}
\end{array} 
\label{bundle} \; ,
\end{equation}
i.e. with the ($d$-1)-dimensional complex projective space. The projection map $\pi : \mathcal{H}_{0} \rightarrow P\mathcal{H}$ is simply given by:
\begin{equation}
\pi : |\psi\rangle \mapsto \frac{|\psi\rangle \langle \psi| }{\langle \psi|\psi\rangle } \; .
\end{equation}
The quotient space $P\mathcal{H}$ is endowed with a K\"ahler structure, given by a metric tensor $g$, the Fubini-Study metric, and a compatible  symplectic structure $\omega$. Together they form a Hermitian structure $h= g + i \omega$, of which they represent  the real and the imaginary part (see \cite{erc} and the aforementioned references for a review). For many purposes it is convenient to represent $h$ directly on $\mathcal{H}_0$  by means of its pull-back via $\pi$:
\begin{equation}
h =  \frac{\langle d\psi |d  \psi \rangle  }{\langle \psi|\psi\rangle } -  \frac{\langle d\psi | \psi \rangle \langle \psi | d\psi \rangle   }{\langle \psi|\psi\rangle ^2} \; ,\label{hermit}
\end{equation}
where here the symbol $d$ denotes differentiation with respect to some chart coordinates in the Hilbert space. Notice that, when fixing the number of parameters and expliciting the $\theta$ dependence, we are restricting the metric tensor to a particular submanifold identified by $\rho(\theta)$ and its lift $\psi(\theta)$.

In this context it has been shown (see, for instance, \cite{marmo}) that the quantum Fisher information  (\ref{qfisher}) (seen as a tensor) can be identified with the Hermitian form (\ref{hermit}). The differential $d\rho$ is interpreted now as a matrix of one forms, i.e. a section of $\Omega^1(\mathfrak{u}(n))$, and straightforward calculations lead to:
\begin{equation}
I = \Tr {\rho  (\dl \rho)^2} = 4 \Tr{\rho (d \rho)^2} = 4 [ \langle d\psi |d  \psi \rangle  - \langle d\psi | \psi \rangle \langle \psi | d\psi \rangle ]  \; , \label{hfisher}
\end{equation}
where we have set $\rho= | \psi \rangle \langle \psi| $ with $\langle \psi|\psi\rangle=1$ and used property (\ref{dd}).\\

We will rephrase now this result in a different way, that will be helpful for the generalization to mixed states that will be discussed in the next section.
Fixing a normalized reference vector $|\psi_0\rangle  \in\H_0$ and acting upon it with the whole unitary $d$-dimensional group $\mathsf{U}(d)$, we are able to reach any other point on the unit sphere $\S^{2d-1}$, with  the stabilizing subgroup being isomorphic to $\mathsf{U}(d-1)$. Hence:
\begin{equation}
\S^{2d-1}=\qsp{\mathsf{U}(d)}{\mathsf{U}(d-1)} \; . 
\end{equation}
To obtain the ray space, we must now quotient out the phase redundancy, getting:\begin{equation}
\label{orbitproj}
\eqalign{
\mathbbm{CP}^{d-1}\sim\P&=\qsp{\left(\qsp{\mathsf{U}(d)}{\mathsf{U}(d-1)}\right)}{\mathsf{U}(1)}\\
&\sim\qsp{\mathsf{U}(d)}{\mathsf{U}(d-1)\times\mathsf{U}(1)} } \; . 
\end{equation}
The base space $\P$ of this principal bundle is (modulo an $i$ factor) a subset of the Lie algebra $\mathfrak{u}(n)$, vector space of all skew-Hermitian matrixes. More specifically we have that the base space is one of the orbits of the co-adjoint action of $\mathsf{U}(d)$ on its Lie algebra (properly on its dual). In fact we may restate the procedure described above as follows: take a reference point $\rho_0$ in $\mathfrak{u}(n)$, $\rho_0 = |\psi_0\rangle \langle \psi_0|$, and act upon it by the (co-)adjoint action of the Lie group:
\begin{equation}
\rho = U \rho_0 U^\dagger \; \; , \;\; U \in \mathsf{U}(d) \; .  \label{rhoref}
\end{equation}
Since $|\psi_0\rangle $ is a fixed vector for the (right) action of the stabilizer subgroup  $\mathsf{U}(d-1)$, the action of this subgroup on $\rho_0$ will be ineffective; moreover $\rho_0$ is $\mathsf{U}(1)$ invariant, so that even the combined action of the two subgroups will leave it unaltered. 

A generic tangent vector $X \in T_\rho \P$ at the point $\rho$, is of the form:
\begin{equation}
X = - i [K,\rho] \;\; , \;\; \mbox{ with } \;\; K^\dagger = K = i (dU)U^\dag = -i U (dU^\dag) \; . 
\end{equation}
If $\rho= |\psi\rangle \langle \psi | $ and $|\chi\rangle $ is a vector orthogonal to $|\psi\rangle$, we have:
\begin{equation}
K = i (|\chi \rangle \langle \psi| - |\psi \rangle \langle \chi|) \;\; , \;\; X = |\chi \rangle \langle \psi| + |\psi \rangle \langle \chi| \;\; \mbox{ with } \;\; \langle \chi | \psi \rangle =0 \; . \label{kappa}
\end{equation}
Given two such vectors $X,X'\in T_\rho \P$, identified by the operators $K,K'$ determined respectively by the two vectors $|\chi\rangle, |\chi '\rangle$, the Fubini-Study metric $G_{FS}$ and the compatible  Kostant Kirillov Souriau (KKS) symplectic form, that together yield the Hermitian structure $ H = G_{FS} + i \Omega_{KKS}$ on the coset space $\P$, are \cite{beng}:
\begin{equation}
\eqalign{  G_{FS}(X,X') &= \frac{1}{2} \Tr{\rho \{K,K'\} } = \mathrm{Re} \langle \chi | \chi' \rangle \\
\Omega_{KKS}(X,X')  &= - \frac{i}{2} \Tr{\rho [K,K' ] } =\mathrm{Im} \langle \chi | \chi' \rangle } \; . 
\end{equation}
The identification of the quantum Fisher information with the real part of such an Hermitian structure follows immediately by noticing that, putting $|d\psi\rangle = i a |\psi\rangle + |\chi\rangle$ with $\langle \chi | \psi \rangle =0$, we have $d\rho = X = -i [K,\rho]$ where $K$ is given as in equation (\ref{kappa}). Indeed we have:
\begin{equation}
I = \Tr {\rho  (\dl \rho)^2} = 4 \Tr{\rho (d \rho)^2} = 2 \Tr{\rho \{K,K\}} \; . \label{ident2}
\end{equation}

To end this section, we would like to present a geometrical  interpretation of formula (\ref{reach}) that gives the condition for the information bound to be attained. In the language we have used in this subsection, condition (\ref{reach}) reads as:
\begin{equation}
\xi_1 \propto_{\mathbb R} \langle \chi | \xi\rangle \label{rcond} \; . 
\end{equation}
Since $ | \chi\rangle$ can always be chosen so that   $\langle \chi | \xi\rangle $ is real,  this means that the vector $|\xi\rangle$ that defines the measure must have real components with respect both to the vector $|\psi\rangle$, which defines a point on the unit sphere, and to $|\chi\rangle$ that defines a tangent vector at this point. This assumes a particular meaning when we consider  the special case of of a two-level system, when $\P=S^2$ since we are looking at the Hopf fibration \cite{bott}:
\begin{equation}
\begin{array}
[c]{rc}%
\mathsf{U}\left(  1\right)   \hookrightarrow & \mathbb{S}^{3}\\
  & \downarrow\\
  & \P \sim \mathbb C \mathbb P^{1} = S^2
\end{array}  \; . 
\label{2bundle}
\end{equation}
In this case (\ref{rcond}) means that $\chi$ belongs to the plane  defined by the unit vector identifying the point of the sphere and its tangent one. This plane  passes through the center of the sphere,  intersects the sphere itself in a great circle and has to be kept fixed if we want the bound to be reached uniformly in the parameter $\theta$. This exactly what was found in \cite{barn,luati1} by using algebraic techniques.

\section{Mixed states for a q-bit} \label{se:mix} 

In this section we will consider a two level mixed quantum system described by the mixing:
 \begin{equation}
\rho=k_1\rho_1+k_2\rho_2 \label{rmix}  \label{mixrho}
\end{equation}
of two pure states $\rho_1=\pure{\psi_1},\ \rho_2=\pure{\psi_2}$, such that $\scalar{\psi_i}{\psi_j}=\delta_{ij}$ and with $k_1,k_2\ge0$, $k_1+k_2=1$. Without loss of generality we may assume $0<k_1\leq1/2 < k_2 < 1$. The limiting cases $k_1=0,\, k_2=1$ and $k_1=k_2=1/2$ correspond respectively  to the case of a pure state ($\rho= \pure{\psi_2}$), discussed in the previous section, and to that of a degenerate mixed state. Throughout the following calculations we will need to exclude these two situations, but we will be able to recover them  at the end of next subsection.

As before, we will assume that $\rho = \rho(\theta)$, $\theta$ being a scalar parameter. We will assume that, as $\theta$ changes, the rank of $\rho$ remains constant, equal to its maximum value two. Clearly the density operator may depend on this parameter through both the value of the constants $k_1(\theta)$ and $k_2(\theta)=1-k_1$ and a variation of the projectors $\rho_1(\theta) = \pure{\psi_1}$ and $\rho_2(\theta) = \pure{\psi_2} = \mathbb I - \rho_1$.  As we will see in the  next  subsection,  these two situations may be first studied separately. We will put them  together in subsect. \ref{se:fishermix} and we will give a geometrical interpretation in subsect. \ref{se:intermix}. We will conclude our analysis in subsect. \ref{se:optmix} by considering an optimization problem similar to the one we have seen for the pure state case.

\subsection{Geometry of  mixed states} \label{se:sphere} 

We will study the geometry of mixed states of the form (\ref{rmix}) first by keeping the weights $k_1$ and $k_2$ fixed. 
It follows directly from the definition that, if we use the ordered pair $\Psi=(\ket{\psi_1},\ket{\psi_2})$ to build  the $2\times 2$ unitary matrix
\begin{equation}
U= \left( \begin{array}{cc} \ket{\psi_1} & \ket{\psi_2} \end{array} \right)   \label{matrixU}
\end{equation}
we may set
\begin{equation}
\rho=U\rho_0U^\dag  \; , \; \mbox{ with } \;\;    \rho_0= \left(\begin{array}{cc}k_1 & 0\\ 0& k_2\end{array}\right) \; .  \label{kdiag}
\end{equation}
Here we are using a representation analogue to that presented in eq. (\ref{rhoref}) for pure states, where now the diagonal matrix $\rho_0$ represents the reference point, on which we act with the group $\mathsf{U}(2)$.  To remove the phase degeneracy of the two vectors $\ket{\psi_j}$, we have to quotient  by the (right) action of the toral subgroup $\mathsf{U}(1)\times \mathsf{U}(1)$ given by matrices of the form:
\begin{equation}
D= \left( \begin{array}{cc} e^{i\phi_1} & 0 \\ 0 & e^{i\phi_2}  \end{array} \right)   \label{matrixD} \; , 
\end{equation}
obtaining the coset
\begin{equation}
\label{coset2}
\eqalign{
\P^{(2)}_2&= \qsp{\mathsf{U}(2)}{\mathsf{U}(1)\times\mathsf{U}(1)}} \; . 
\end{equation}
Like in the case of a two level pure state, this is again  the two-dimensional sphere $S^2$. We remark however that the principal fiber bundle structure we are considering is different: instead of the fibration (\ref{2bundle}) we now have \cite{chat} 
\begin{equation}
\begin{array}
[c]{rc}%
\mathsf{U}(1) \times  \mathsf{U}(1) \hookrightarrow & \mathsf{U}(2)\\
  & \downarrow\\
  & \P_2^{(2)} \sim  S^2
\end{array} \; . 
\label{mixbundle}
\end{equation}

Points in the coset space, which is an example of a flag manifold \cite{adel,pick},  may be represented for instance by means of the following explicit parametrization:
\begin{equation}
U=U(z)=\frac{1}{\sqrt{1+|z|^2}}\left(\begin{array}{cc}
|z| & e^{i\chi} \\
-e^{-i\chi} & |z|
\end{array}\right) \label{stereo} \; , 
\end{equation}
where $z=|z|e^{i\chi}$, which corresponds to stereographic coordinates on the sphere, and will be useful in what follows. Notice that, in writing this expression, we have made an explicit choice of the phases of the two vectors $\ket{\psi_j}$'s, since the abelian subgroup (\ref{matrixD}) acts by:
\begin{equation}
U(z) \mapsto U(z) D =\frac{1}{\sqrt{1+|z|^2}}\left(\begin{array}{cc}
|z| e^{i\phi_1} & e^{i\chi}e^{i\phi_2}  \\
-e^{-i\chi} e^{i\phi_1} & |z| e^{i\phi_2} 
\end{array}\right) \; . 
\end{equation}
Thus, expression (\ref{stereo}) may be seen as a particular lift of a point of the two-dimensional sphere in the total space $\mathsf{U}(2)$, corresponding to the choice $\phi_{1}=\phi_{2} =0$, but this has no consequences in the following calculations. With this parametrization, the density matrix reads as:
\begin{equation}
\rho(z)= U(z) \rho_0 U^\dag(z) = \frac{1}{1+|z|^2}\left(\begin{array}{cc} 
k_1|z|^2+k_2 & (k_2-k_1)z \\ 
(k_2-k_1)z^* & k_1+|z|^2k_2
\end{array}\right) \; . 
\end{equation}  

Now, $\P^{(2)}_2\sim S^2$ is a K\"ahler manifold endowed with an Hermitian structure that, in the  local chart we are using, reads \cite{pick}:
\begin{equation}
h=\frac{4}{(1+|z|^2)^2}\left(\d z\odot\d z^* + i \,  \d z\wedge\d z^*\right) \; , 
\end{equation} 
 where $dz \odot dz^* =  (dz\otimes dz^* + dz^*\otimes dz)/2, \; dz\wedge dz^*=(dz\otimes dz^* - dz^*\otimes dz)/2i$ represent the symmetric and antisymmetric tensor product respectively, yielding\footnote[7]{If we use spherical coordinates, such that $z=cotan(\theta/2) e^{i\phi}$,  we see that the above expressions correspond to the standard metric and volume form on the sphere: $g= (d\theta)^2+\sin^2\theta\, (d\phi)^2$ and $\omega=\sin\theta\, d\theta\wedge d\phi$.} the FS-metric:
\begin{equation}
g \equiv \frac{4}{(1+|z|^2)^2}\, \d z\odot\d z^*  \label{gs}
\end{equation}
and the KKS-simplectic form:
\begin{equation}
\omega \equiv \frac{4 }{(1+|z|^2)^2}\, \d z\wedge \d z^* \label{omegas} \; . 
\end{equation} 
We remark that, here and in the following, the symbol $d$ to denote the proper De Rham differential on differential forms, with values in the Lie algebra, with $dz, dz^*$ representing the one-forms associated to the (complex) chart coordinates we have chosen to describe the space of states.

~\linebreak

Up to now, we have kept the coefficients $k_1$ and $k_2=1-k_1$ fixed and seen that the corresponding space of states is homeomorphic to a two-dimensional sphere $S^2$: this holds for both the pure case ($k_1=0$) and for the mixed state situation ($0<k_1\leq 1/2$). We may easily conclude \cite{zyc} that the space of rank-2 mixed states for a two-level systems is homeomorphic to $S^2 \times [0,1/2]$. Thus, when working on the sphere, we have at our disposal all the geometric structures defined above. Instead, the transverse direction is obtained by assuming that in expression (\ref{mixrho}) the vectors $\ket{\psi_j}$'s are kept fixed, and it is simply given by a segment whose boundaries correspond to the pure state case and the maximally mixed one. 

\subsection{Fisher metric for mixed states}   \label{se:fishermix} 

Let us first analyze the case in which the weights $k_j$ are kept fixed, meaning that we are working on a sphere. A straightforward calculation starting from  (\ref{stereo}) gives the following expression for the one form:
\begin{equation}
\d\rho=\frac{(k_1-k_2)}{(1+|z|^2)^2}\left(\begin{array}{cc} 
z^*\d z+z\d z^* & z^2\d z^*-\d z \\ 
{z^*}^2\d z-\d z^* & -z^*\d z-z\d z^*
\end{array}\right) \; . 
\end{equation}  
Notice that $\d\rho$ is the differential of a Hermitian matrix with unitary trace, it has therefore a particular simple expression, namely
$$\left(\begin{array}{cc}A&B \\ B^*& -A\end{array}\right) \; , $$ 
so that its square is a multiple of identity, provided that we define the square operation by means of a symmetrized tensor product, that is $\d\rho^2\equiv\d\rho\odot\d\rho$. Performing the calculation explicitly we find:
\begin{equation}
\label{RHO}
(\d\rho)^2=\d\rho\odot\d\rho=\frac{(k_1-k_2)^2}{(1+|z|^2)^2}\mathbb I \, \d z\odot\d z^* \; . 
\end{equation}

Now, using the fact $\rho_1 = \mathbb I - \rho_2$ and thus that $d\rho_1 = - d\rho_2$ (where $\rho_j$ are rank-one projectors satisfying all properties seen in the previous section),  it is not hard to verify that \cite{luati1}:
\begin{equation}
\rho d\rho + d\rho \rho = (k_1+ k_2) d\rho \; .\end{equation}
Hence the symmetric logarithmic one form $\dl\rho$, which is a section of $\Omega^{(1)}(\mathfrak{u}(2))$ as $d\rho$ and is implicitly defined through eq. (\ref{logder}), is given by\footnote[7]{Here we have kept the denominator, even if it equals one,  to make the symmetric logarithmic differential an adimensional quantity and preserve the correct homogeneity in equation \eref{logder}.}:
\begin{equation}
\dl\rho=\frac{2}{k_1+k_2}\d\rho= 2\d\rho \; . 
\end{equation}

Similiarly to the pure state case, we may prove that the quantum Fisher metric (\ref{qfisher}) is given by:
\begin{equation}
I_\theta=\Tr{\rho(\dl\rho)^2} = 2 \Tr{(\d\rho)^2}  = \frac{4(k_1-k_2)^2}{(1+|z|^2)^2} \, \d z\odot\d z^* \; , \label{irho}
\end{equation}
where we have used formula (\ref{RHO}). This means that the Fisher information metric is proportional to the Fubini-Study metric, the proportionality constant depending on the specific mixing coefficients.

We notice  that if we fix, say, $k_i=1$, it is easy to see that 
\begin{equation}
\label{drhoi}
4\Tr{ \rho_i (\d\rho_i)^2}=\frac{4}{(1+|z|^2)^2} \, \d z\odot\d z^*
\end{equation}
and therefore, comparing \eref{irho} and \eref{drhoi}, one is able to show directly another of the results found in \cite{luati2}, namely that
\begin{equation}
I_\theta =(k_1-k_2)^2 I_i  \; , 
\end{equation}
where $I_i\equiv\Tr{\rho_i(\dl\rho_i)^2}$, which means that the Fisher information index when mixing occurs is always less than the one  of a pure state.\\

Let us now consider the transverse direction to the sphere, by allowing only the value of $k_1=k$ (and $k_2=1-k$) to change smoothly.
When $\pure{\psi_1}$ and $\pure{\psi_2}$ are kept constant, for any value of the parameter $\theta$ the matrix $\rho$ may be cast in the diagonal form
\begin{equation}
\rho(\theta) = \left( \begin{array}{cc} k(\theta) & 0 \\ 0 & 1-k(\theta) \end{array} \right) \; , 
\end{equation}
so that the logarithmic form is simply given by:
\begin{equation}
\dl\rho(\theta) = \d \left( \begin{array}{cc} \log k & 0 \\ 0 & log(1-k) \end{array} \right) = 
\frac{\d k}{k}\rho_1-\frac{\d k}{1-k}\rho_2
\; , 
\end{equation}
which can be easily calculated:
\begin{equation}
\label{deellerhotrans}
\dl\rho=\frac{ \d k }{(1+|z|^2)k(1-k)}\left(\begin{array}{cc}
|z|^2-k(|z|^2+1) & -z \\
-z^* & 1-k(|z|^2+1)
\end{array}\right) \; . 
\end{equation}
From the definition (\ref{qfisher}), it is just a long but straightforward computation to obtain that the quantum Fisher metric is now given by:
\begin{equation}
I = \label{luatifisherlink2}
\Tr{\rho\,(\dl\rho^2)}=\frac{\d k\odot\d k}{k(1-k)} \; . 
\end{equation} 
Let us notice that the logarithmic derivative along the transverse direction is generated by the diagonal matrix $\sigma_z$, while on the sphere tangent directions were generated by the off-diagonal elements $\sigma_{\pm}=\sigma_x \pm i \sigma_y$. We will come back to this point in the next subsection.

\subsection{Geometric interpretation for mixed states} \label{se:intermix} 

Putting the different results obtained in the previous subsection together, we conclude that the quantum Fisher metric for a mixed state of a q-bit is given by:
\begin{equation}
I_{tot} = \frac{\d k\odot\d k}{k(1-k)} +  \frac{4(k_1-k_2)^2}{(1+|z|^2)^2} \, \d z\odot\d z^* \; , \label{qftot}
\end{equation}
a result in agreement with what found in \cite{luati2}.
Recalling expression (\ref{gs}) for the metric on a unit sphere in stereographic coordinates, we may interpret the second term as  the usual metric on a sphere of radius  $r= k_1-k_2 = 1-2k_1$.  It is worth noticing that the first term does not depend on the point on the sphere, suggesting that the global metric on the space of all possible mixed states has to be spherically symmetric. Also, if we now take the segment $k\in ]0,1/2[$ parametrized by the radius $r=1-2k\in ]0,1[$, equation (\ref{luatifisherlink2}) corresponds to the following metric $g_t$ along the transverse direction:
\begin{equation}
g_t = \frac{ \d r\odot\d r}{1-r^2} \label{metricr}  \; . 
\end{equation}
By making the simple change of variable $r=\sin \Psi$, with $\Psi\in ]0,\pi/2[$, we see that: $g_t =  d\Psi \odot d\Psi $. Thus we may interpret the transverse direction as a quarter of a unit circle, on which we calculate distances as arclengths. We also see that the singularity for $k=0$ or $r=1$ appearing in the expression for the metric (\ref{luatifisherlink2}) when reaching the boundary of the transverse direction  is just an artifact of the coordinates chosen. This boundary corresponds to the case where $\rho$ represents a pure state, which we may therefore correctly recover as a limit case of the most general mixed state situation. 
On the opposite boundary ($k=1/2$ or $r=0$), the transverse direction is well defined, but the coefficient in front of the second term vanishes ($r^2=0$). This is what is expected if we recall that, when $k_1=k_2=1/2$,  the bundle (\ref{mixbundle}) has to be changed. In this case the density matrix $\rho$ is degenerate (in fact, it is a multiple of the identity) and thus invariant under the full group $\mathsf{U}(2)$: the base space is $\mathsf{U}(2)/ \mathsf{U}(2)$ given then by just a single point, that may be seen as a sphere of null radius.

It is well known that different metrics may be defined on the space of mixed states and used for different purposes (see \cite{zyc} for a review). Thus, it is interesting to notice that formula (\ref{qftot}) represents the standard metric on the three-dimensional unit sphere $S^3= \{(x_1,x_2,x_3,x_4) \, : \, \sum_{j=1}^4 (x_j)^2 =1 \}$. Indeed, by setting:
\begin{equation}
\eqalign{ &x_1 = (k_1 - k_2 ) \sin\Psi \sin \theta \cos \phi =   \frac{2(k_1-k_2)}{\sqrt{1+|z|^2}} \sin\Psi Re(z)\\
&x_2 = (k_1 - k_2 ) \sin\Psi \sin \theta \sin \phi =  \frac{2(k_1-k_2)}{\sqrt{1+|z|^2}}  \sin\Psi Im(z)\\
& x_3 =(k_1 - k_2 ) \sin\Psi \cos \theta =  \sin\Psi \frac{|z|^2-1}{1+|z|^2} \\
& x_4 = \cos\Psi } \; \; \; , 
\end{equation}
one finds that (\ref{qftot}) can be written as:
\begin{equation}
g_{S^3} = (d\Psi)^2+ \sin^2\Psi \left[ (d\theta)^2 + \sin^2\theta (d\phi)^2 \right] = I_{tot} \; . 
\end{equation}
This result has also an important interpretation in terms of Lie groups. Indeed, this shows that the Fisher metric on the space of states is just the (equivariant) metric on the space $S^3= \mathsf{S}\mathsf{U}(2)$ that can be seen as obtained  from the full group $\mathsf{U}(2)$ by quotienting out the trivial action of the subspace $\mathsf{U}(1)= \{ e^{i\alpha \mathbb I}\}$. The quotient of $ \mathsf{S}\mathsf{U}(2)$ with respect to the action of its abelian subgroup $\mathsf{U}(1)= \{ e^{i\alpha' \sigma_z}\}$ yields the two-dimensional sphere of mixed states with fixed rank (equal to 2) and $(k_1,k_2)$, while changing the latter corresponds to moving along the fiber. \\

We shall present now an alternative derivation of the fact that, when keeping $k_1,k_2$ constant,  the quantum Fisher information is given by the Fubini-Study metric on the coset space $S^2$, by following a calculation similar to the one used in the second part of subsect. \ref{interpure}.
If $\rho= k_1 \pure{\psi_1} + k_2 \pure{\psi_2}$, we set:
\begin{equation}
\eqalign{ & \ket{\d \psi_1} = ia \ket{\psi_1} + \lambda \ket{\psi_2}  \\
& \ket{\d \psi_2} = - \lambda^* \ket{\psi_1} + i b  \ket{\psi_2}  } \;\; , 
\end{equation}
with $a,b\in \mathbb R$ and $\lambda \in \mathbb C$, as it follows form the conditions $\langle \psi_i | \psi_j \rangle = \delta_{ij}$ that imply $\langle \psi_i | \d \psi_i \rangle = - \langle \psi_i | \d \psi_i \rangle^*$   and   $ \langle \psi_1 | \d \psi_2 \rangle = - \langle   \psi_2 | \d \psi_1 \rangle^*$. Then, it is not difficult to gather that the expressions for $\d \rho$ and $K$ such that $\d \rho = -i [K,\rho]$ are given by:
\begin{equation}
\eqalign{ & \d \rho = (k_1-k_2) \left( \lambda^* \ket{\psi_1} \bra{\psi_2} +  \lambda \ket{\psi_2} \bra{\psi_1} \right)\\
&K =  -i \left( \lambda^* \ket{\psi_1} \bra{\psi_2} -  \lambda \ket{\psi_2} \bra{\psi_1} \right) \label{kappamix} } \; . 
\end{equation}
For a generic $\lambda\in \mathbb C$, the Hermitian matrix $K$ given in (\ref{kappamix}) represents the generator of the co-adjoint action corresponding to the tangent vector $X= -i [K, \rho]\in T_\rho \P^{(2)}$, so that we have:
\begin{equation}
G_{FS}(X,X') = \frac{1}{2} \Tr{\rho \{K,K'\} } = \mathrm{Re} \left[ \lambda^* \lambda'\right] \; . 
\end{equation}
In particular $ G_{FS}(d\rho,d\rho)= |\lambda|^2$. A straightforward, still lengthy, calculation then shows that:
\begin{equation}
\Tr{\rho (\d \rho)^2 } = (k_1-k_2)^2  |\lambda|^2 = (k_1-k_2)^2 G_{FS}(\d\rho,\d\rho) \; , 
\end{equation}
which yields the desired identification of the quantum Fisher information with the Fubini-Study metric.

Let us remark that for the two level mixed states case the Fisher information metric turns out to be again the Fubini-Study metric on the sphere just because the base space of the fibration is the projective space both in the pure and mixed state case.
As a matter of fact, if one considers more complicated mixings the space of states differs sensibly: for example with $d=3$ one has to consider the flag manifold  \cite{adel}
\begin{equation}
\mathcal{P}^{3}_3\equiv\mathbbm{F}_3\simeq \qsp{U(3)}{U(1)^3}
\end{equation}
which is not projective. The fact that the Fisher information metric is somehow naturally related to the metric that the space of states can be endowed with is still an open question in the general case. We hope that the general geometric framework we are presenting might clarify this point.

\subsection{The optimization problem for mixed states}    \label{se:optmix} 

In this section we conclude our discussion of mixed states for a q-bit by looking at the optimization problem. This case has already been treated and solved in \cite{luati2}: here we will simply rephrase the arguments in our language.

We want to write the general formula (\ref{condition}) that gives the optimization condition. If, as before, we consider only projective measures, we may assume:
\begin{equation}
m^{1/2} = c \ket{\gamma} \bra{\gamma} \;\; \mbox{with} \;\; \langle\gamma|\gamma\rangle=1 \;\; c>0 \; \Rightarrow m = \ket{\xi} \bra{\xi} \; , 
\end{equation}
where $\ket{\xi} = c \ket{\gamma}$. In the basis $\{\ket{\psi_1}, \ket{\psi_2}\}$, with respect to which $\ket{\xi} = \xi_1 \ket{\psi_1} + \xi_2 \ket{\psi_2}$, we can write:
\begin{equation}
m^{1/2} = \frac{1}{c} \left( \begin{array}{cc} |\xi_1|^2 & \xi_1 \xi_2^* \\ \xi_1^* \xi_2 & |\xi_2|^2 \end{array} \right) \; . 
\end{equation}
Also, in this basis, we have:
\begin{equation}
 \rho^{1/2} = \left( \begin{array}{cc} \sqrt{k_1} & 0 \\ 0& \sqrt{k_2}  \end{array} \right) 
\end{equation}
and
\begin{equation}
\dl  \rho = \left( \begin{array}{cc}1/k_1 & 2 (k_1-k_2) \lambda^* \\  2 (k_1-k_2) \lambda & 1/k_2  \end{array} \right) \; . 
\end{equation}
Thus, equation (\ref{condition}) translates into a set of four conditions:
\begin{equation} \eqalign{
& |\xi_1|^2 = R \left[ \frac{|\xi_1|^2}{k_1} + 2 (k_1-k_2) \lambda \xi_1 \xi_2^* \right] \\
& |\xi_2|^2 = R \left[ -\frac{|\xi_2|^2}{k_2} + 2 (k_1-k_2) \lambda^* \xi_1^* \xi_2\right] \\
& \xi_1^* \xi_2 = R \left[ \frac{\xi_1^* \xi_2}{k_1} + 2 (k_1-k_2) \lambda |\xi_2|^2 \right] \\
& \xi_1 \xi_2^* = R \left[ -\frac{\xi_1\xi_2^*}{k_2} + 2 (k_1-k_2) \lambda^* |\xi_1|^2\right] } \;\;\;\;\; , \mbox{ with } R\in\mathbb R\; . \label{condmixed}\
\end{equation}
It is immediate to see that these equations imply:
\begin{equation}
\lambda \xi_1 \xi_2^* \in \mathbb R
\end{equation}
and that the coefficients $\xi_1,\xi_2$ may be chosen to be real proportional. Thus the situation is similar to what found in the pure state case.
We refer to \cite{luati2}  for further details about the calculations.

\section{Geometric meaning of the Fisher tensor} \label{se:geofisher} 

\subsection{The Fisher tensor} \label{se:fishertensor} 
In what we did until now, $\rho(\theta)$ was understood to be that particular (real) one-dimensional submanifold of $S^2$ determined by the dependence from the single parameter $\theta$. Thus, in the expression for $I(\theta)=\Tr{\rho(\dl\rho)^2}$ we always meant to take the square as a symmetrized tensor product, and this assumption was justified by the fact that any possible antisymmetric part of the tensor product would have vanished on the one dimensional curve $\rho(\theta)$ (More precisely it would vanish after the pullback on the parameter space, which is the one dimensional line). 
Now we would like to relax this restriction and compute instead the element $\mathfrak{F}\in(T^*S^2)^{\otimes 2}$, the full Fisher tensor, defined via:
\begin{equation}
\label{fishertensor}
\mathfrak{F}=\Tr{\rho\dl\rho\otimes\dl\rho} \; . 
\end{equation}

Starting from (\ref{stereo}), it is  just a matter of computations and smart rewritings to see that:
\begin{equation}
\eqalign{
\fl\dl\rho\otimes\dl\rho=4\d\rho\otimes\d\rho=\frac{4(k_1-k_2)^2}{(1+|z|^2)^2}\left[\mathbbm{I}\d z\odot\d z^*+i\left(\begin{array}{cc}
|z|^2-1 & -2z \\
-2z^* & 1-|z|^2
\end{array}\right)\frac{\d z^*\wedge\d z}{(1+|z|^2)}\right] 
} \; .
\end{equation} 
Notice that the matrix coefficient of the antisymmetric tensor $\d z^*\wedge \d z$ is just the adjoint transformed of the Lie algebra generator $\sigma_z$:
\begin{equation}
U(z)\sigma_zU^\dag(z)=\frac{1}{(1+|z|^2)}\left(\begin{array}{cc}
|z|^2-1 & -2z \\
-2z^* & 1-|z|^2
\end{array}\right) \; . 
\end{equation}
while the symmetric part can be seen as trivially multiplied by $\mathbbm{I}=U(z)\mathbbm{I}U^\dag(z)$. This implies that we can write 
the complete Fisher tensor as:
\begin{equation} \label{explicitfishertensor}
\eqalign{
\mathfrak{F}&=\frac{4(k_1-k_2)^2}{(1+|z|^2)^2}\Tr{\rho(U(z)\mathbbm{I}U^\dag(z)\d z\odot\d z^* + iU\sigma_zU^\dag\d z^*\wedge\d z)}=\\
&=\frac{4(k_1-k_2)^2}{(1+|z|^2)^2}\left[ (k_1+k_2)\d z\odot\d z^*-i(k_1-k_2)\d z\wedge\d z^* \right]
}
\end{equation}
since, for instance, $\Tr{\rho U\sigma_z U^\dag}=\Tr{\rho_0\sigma_z}=(k_1-k_2)$.

It is worthwhile to see another equivalent way to compute the same tensor, because it will give us some important insights. From the definition of $d\rho$ and $U(z)$, it is easy to see that we may write:
\begin{equation}
\eqalign{
d\rho=U(z)d\rho_0U^\dag(z)   \;\; \mbox{ with}\\
d\rho_0=-i[K_0,\rho_0] \;\; , \;\;  K_0 = i \, U^\dagger dU \; .
}\end{equation}
Now, $K_0$ and $d\rho_0$ necessarily have the following matrix structure:
\begin{equation}
\eqalign{
K_0=i\left\{\left(\begin{array}{cc}0 &  0 \\ \lambda & 0 \end{array}\right)dz-\left(\begin{array}{cc}0  &  \lambda^* \\ 0 & 0 \end{array}\right)dz^* \right\} \\
d\rho_0=(k_1-k_2)\left\{\left(\begin{array}{cc}0 &  0 \\ \lambda & 0 \end{array}\right)dz+\left(\begin{array}{cc}0 &  \lambda^* \\ 0 & 0 \end{array}\right)dz^*\right\}
} \; , \end{equation}
as it can be seen by direct inspection, with:
\begin{equation}
\lambda = \frac{e^{-2i\chi}}{(1+|z|^2)} \; . 
\end{equation}
Rewriting $\mathfrak{F}$, it is then just a straightforward computation to see that 
\begin{equation}
\mathfrak{F}=\Tr{\rho(2d\rho)\otimes(2d\rho)}=4\Tr{\rho_0d\rho_0\otimes d\rho_0}
\end{equation}
agrees with \eref{explicitfishertensor}. This formula is particularly interesting, because it explicitly shows that the Fisher tensor is equivariant, in the sense that its value can always be calculated at the fiducial point $\rho_0$\footnote[7]{Notice that this rewriting makes also evident the relation between the $(z,z^*)$ and the $(\ket{\psi_1},\ket{\psi_2})$ parameterizations.}.

Thus, given two generic vectors at  $\rho_0$:
\begin{equation}
X_0(v)=v\frac{\partial}{\partial {z}}+v^*\frac{\partial}{\partial {z^*}} \;\; , \;\; X_0(v')=v'\frac{\partial}{\partial {z}}+v'^*\frac{\partial}{\partial {z^*}} 
\end{equation}
with $v,v'\in \mathbb C$, one finds:
\begin{equation}\label{fishertensorcomputed}
\fl\mathfrak{F}(X_0(v),X_0(v'))=\frac{4(k_1-k_2)^2}{(1+|z|^2)^2}\left((k_1+k_2)\mathrm{Re}(v^*v')+i(k_1-k_2)\mathrm{Im}(v^*v')\right)
\; . \end{equation}

\subsection{Relation to the KKS construction} \label{se:kksintermix} 

Exploiting the co-adjoint action of a Lie group, say $G$, on its dual Lie algebra $\mathfrak{g}^*$ it is possible to show that one gets a symplectic foliation of $\mathfrak{g}^*$, each leaf corresponding to the orbit ${\cal O}(\rho_0)=\left\{ \rho= U \rho_0 U^{-1} \; : U\in G\right\}$ of a particular reference point $\rho_0\in\mathfrak{g}^*$. The natural symplectic structure that renders each orbit a symplectic manifold is the so called Kostant-Kirillov-Souriau form $\Omega_{KKS}$ \cite{kir}. 

The map $K_0 \mapsto \cad_{K_0}\rho_0$ identifies $\mathfrak{g}/\mathfrak{g}_0$ with the tangent space to ${\cal O}(\rho_0)$ at the point $\rho_0$, where $\mathfrak{g}_0$ is the Lie subalgebra of the stabilizer subgroup for $\rho_0$. A generic tangent vector at $\rho= U \rho_0 U^{-1} $ is of the form: 
\begin{equation}
X_\rho = \cad_{K}\rho =-i[K, \rho] = U \, \cad_{K_0}\rho_0\, U^{-1},\ \ \ \ K=UK_0U^{-1}
\end{equation}
Then, on the orbit, one can define the KKS symplectic form:
\begin{equation}\label{kkssymp}
\Omega_{KKS} (X_\rho, X'_\rho)=-\frac{i}{2}\Tr{\rho\left[K,K'\right]} \; . 
\end{equation}
For the case we are considering, where $G=\mathsf{U}(2)$ and $Stab(\rho_0) = \mathsf{U}(1)\times \mathsf{U}(1)$  for which ${\cal O}(\rho_0)  = S^2$, it is well known that $\Omega_{KKS}$ is just the standard volume form on the two-dimensional unit sphere \cite{pick}. Thus we may conclude that the imaginary part of the Fisher tensor we calculated in the previous subsection is just proportional to $\Omega_{KKS}$ via the square of the radius of the orbit, $r=|k_1-k_2|$:
\begin{equation}
\mathrm{Im} \, \mathfrak{F}(\cdot,\cdot) = 4 r^2 \, \Omega_{KKS} (\cdot, \cdot) \; . 
\end{equation}
The factor $4$ arises from the fact that the Fisher tensor was calculated on $\dl \rho = 2 d\rho$. 
We would like to recall that the KKS two-form is important also to define and study  Berry's phase for mixed states, as it was demonstrated in \cite{chat}, where the same bundle structure of subsect. \ref{se:sphere} was used, but with a different parametrization.

There is a dual picture of the construction presented in the previous subsection that provides an interesting point of view. The curve 
\begin{equation}
\rho(t) = U(z(t)) \rho_0 U(z(t)) ^\dagger
\end{equation}
has tangent vector at $\rho(t=0) = \rho$ that may be written as:
\begin{equation}
X_\rho(v) = U(z) X_{\rho_0} U(z)^\dagger \; , 
\end{equation}
where $X_{\rho_0}$ is the vector in $T_{\rho_0}S^2$ given by:
\begin{equation}
X_{\rho_0}(v)=(k_1-k_2) \left\{ \left( \begin{array}{cc}0  &  0 \\ v\lambda & 0 \end{array}\right) \frac{\partial}{\partial z} +  \left(
\begin{array}{cc}0 &  v^*\lambda^* \\ 0 & 0 \end{array}\right) \frac{\partial}{\partial z^*} \right\}
\end{equation}
with $(v,v^*) = (dz/dt,dz^*/dt)|_{t=0}$. Now, taking just the matrix part of this vector:
\begin{equation} \label{matx}
\widetilde{X}_0(v) = (k_1-k_2)  \left( \begin{array}{cc}0  &  v^* \lambda^* \\ v\lambda & 0 \end{array} \right) \; , 
\end{equation}
 one finds:
\begin{equation}
\frac{1}{4}\mathrm{Im} \, \mathfrak{F}(X_\rho(v), X_\rho(v')) = -\frac{i}{2}\Tr{\rho\left[\widetilde{X}_\rho(v),\widetilde{X}_\rho(v')\right]}\; . 
\end{equation}
To show this, one may notice that if this equality holds for $\rho_0$, it's easy to see that the extension to any $\rho$ holds as well, due to the invariance of the tensors under the action of $\Ad_{U(z)}$. Thus, it is just necessary to prove that the equality
holds for the reference point, a fact that can be directly checked with the help of \eref{matx}. \\

We would like to see if it is possible to give a complete description of the Fisher tensor in terms of the KKS construction, we shall consider then the following definition of a metric tensor on the co-adjoint orbit. This is related to the fact that on the orbit itself it is possible to define a complex structure $J$ \cite{helg} and a compatible \cite{erc} metric tensor, which makes the orbit a K\"ahler manifold. Such compatible metric is given by:
\begin{equation}
G_{KKS}(\cdot,\cdot)= \Omega_{KKS}(\cdot,J(\cdot))\; . 
\end{equation}
In the complex coordinates $(z,z^*)$ we are using, the complex structure $J$ may be derived just from the change of coordinates $(z,z^*) \mapsto (w=iz,w^*=-iz^*)$. We skip the calculations and only give the final result:
\begin{equation}
J(\widetilde{X}_0(v) ) = \widetilde{X}_0(iv)= (k_1-k_2)  \left( \begin{array}{cc}0  &  -iv^* \lambda^* \\ iv\lambda & 0 \end{array} \right) \; . 
\end{equation}
From this formula, it is easy to find that :
\begin{equation}\label{propdue}
\fl G_{KKS}(\widetilde{X}_\rho(v),\widetilde{X}_\rho(v'))\equiv  \Omega_{KKS}(\widetilde{X}_\rho(v),J(\widetilde{X}_\rho(v'))) =  (k_1-k_2)^3 |\lambda|^2 Re(v^*v')  \; . 
\end{equation}
Comparing with (\ref{fishertensorcomputed}), we see that $G_{KKS}$ is only proportional to the real part of the quantum Fisher tensor, up to a factor of $4r$. This is an interesting consideration that might become important when studying the same problem for more general situations, in which the co-adjoint orbit is not simply a sphere.\\

Finally, we observe that it is not difficult to check that:
\begin{equation}
\frac{1}{4}\mathrm{Re} \, \mathfrak{F}(X_\rho(v), X_\rho(v')) =  \frac{1}{2}\Tr{ \rho\left\{\widetilde{X}_\rho(v),\widetilde{X}_\rho(v')\right\}} \; , 
\end{equation} 
where $\{,\}$ denotes the anti-commutator bracket, and $\widetilde{X}_\rho(v),\widetilde{X}_\rho(v')$ are again the matrices corresponding to the tangent vectors $X_\rho, X'_\rho$ to the co-adjoint orbit over the point $\rho$. Thus, summing up,  we have shown that the full Fisher tensor $\mathfrak{F}$, which is  is manifestly equivariant under the action of the Lie group $U(2)$, can be calculated as: 
\begin{equation}
\fl \frac{1}{4}\mathfrak{F}(X(v),X(v'))= \frac{1}{2}\left\{ \Tr{ \rho\left\{\widetilde{X}_\rho(v),\widetilde{X}_\rho(v')\right\} } + \Tr{\rho\left[\widetilde{X}_\rho(v),\widetilde{X}_\rho(v')\right]} \right\} \; . 
\end{equation}

\section{Conclusions} \label{se:concl}

In this paper we have discussed within a purely geometrical approach the meaning of the Quantum Fisher index and reconsidered the problem of optimization of the Classical Fisher information index. We have first re-derived the pure state case of a q-dit, extending the result of \cite{barn} to any dimension and making contact with \cite{marmo}, to move then to consider the case of a mixed state for 2-level systems, i.e. for q-bits. 

More specifically, in the latter case we have described the geometrical structure of the space of mixed states and shown that it is homeomorphic to $S^2 \times [0,1/2]$. The two-dimensional sphere $S^2$ represents the submanifold of density matrices with fixed maximum rank and eigenvalues $(k,1-k)$, while the interval yields the transverse direction along which we allow for variations in $k$. 

We have also seen that $S^2$ may be interpreted as the the base space of the principal fiber bundle $\mathsf{U}(2)/\mathsf{U}(1)\times \mathsf{U}(1) \sim \mathsf{S}\mathsf{U}(2)/\mathsf{U}(1)$. This geometrical description allows for an identification of the quantum Fisher metric with the standard Fubini-Study metric that one can put on the two-dimensional sphere, as it happens for pure states of any dimension. Also, we have shown that, after including the transverse direction,  the total quantum Fisher metric is nothing but the equivariant standard metric on $\mathsf{S}\mathsf{U}(2)\sim S^3$. 

In the last section, we have introduced the new notion of Fisher tensor and discussed the relation of its imaginary part  with the KKS  symplectic structure that can be defined on co-adjoint orbits of Lie algebras, its symmetric/real part being the usual Fisher information metric. We have also discussed the role of the complex structure.

The case of rank-2 mixed-states for a q-bit is the simplest example one can develop. For it, explicit calculations of the quantum Fisher metric are possible, since one can easily find the symmetric logarithmic derivative starting form the very definition of it. This kind of algebraic calculations become much harder for higher-dimensional Hilbert spaces, i.e. for q-dits with $d\geq 3$. On the contrary, a geometrical approach is still possible. Indeed, for instance, the manifold of fixed-rank (equal to 2) density matrices for q-dits is obtainable as the coset space:
\begin{equation}
\label{coset3}
\eqalign{
\P^{(2)}_d&=\qsp{\left(\qsp{\mathsf{U}(d)}{\mathsf{U}(d-2)}\right)}{\mathsf{U}(1)\times\mathsf{U}(1)}\\
&\sim \qsp{\mathsf{U}(d)}{\mathsf{U}(d-2)\times\mathsf{U}(1)\times\mathsf{U}(1)}} \; \;\;\; . 
\end{equation}

Of course, the geometry of the space of states increases in complexity both with the dimension of the Hilbert space and with the rank of the density matrices. However, the structures we have discussed (fiber bundles, compatible metrics, symplectic forms and so on) are general features of the spaces that can be obtained as co-adjoint orbits of (semi-simple) Lie groups. 
Thus, we are confident that the geometrical interpretation we have discussed in this paper may shed some light on the meaning and properties of the quantum Fisher tensor also in more general systems. Studies in this direction are under consideration.

\section*{Acknowledgments}

We wish to thank Giuseppe Marmo and Giuseppe Morandi for interesting discussions. E.E. thanks also R. Kulkarni for having pointed out some questions addressed in this paper. E.E. was supported in part by  INFN COM4 grant NA41. M.S.  acknowledges partial support of SNF Grant No. PDFMP2\_137103/1.

\end{document}